\documentclass{aa}
\usepackage{psfig}

\begin{document}

%%%%%%%%%%%%%% MY DEFINITIONS
\def\pdot {\dot P}
\def\Omdot {\dot \Omega}
\def\ltsima{$\; \buildrel < \over \sim \;$}
\def\lsim{\lower.5ex\hbox{\ltsima}}
\def\gtsima{$\; \buildrel > \over \sim \;$}
\def\gsim{\lower.5ex\hbox{\gtsima}}
\def\msole{~M_{\odot}}
\def\mdot {\dot M}
\def\gr {GRB030329~}
\def\xte {\textit{Rossi-XTE~}}
\def\xmm  {\textit{XMM-Newton~}}
%%%%%%%%%%%%%%%%%%%%%%%%%%%%%%%%%%%%%%

%   V0.1 - SM - 24/05/2003
%   V0.3 - 27/05   - con discussion di GG+EMR
%   V0.4 - SM figures, abstract, last conclusions, etc..
%   V0.5 - AT revision before submission
%   V1.0 - SM  19-20/06   after referee report. New XMM obs. added
%   V1.2 - SM 24/06   changed order of presentation
%   V1.3 - SM+GG+ER 25/06
%   V1.4 - SM new fig with two XMM lcurves. few other minor changes
%   V2.0 - AT final version

% \special{!userdict begin /bop-hook{gsave 150 90 translate
% 55 rotate /Times-Roman findfont 60 scalefont setfont
% 0 0 moveto 0.5 setgray (revised draft 1.3 - 25/6/03) show grestore}def end}

\title{The X-ray afterglow of GRB030329}

   \author{A. Tiengo\inst{1}$^{,}$\inst{2},
   S. Mereghetti\inst{1},
   G. Ghisellini\inst{3},
   E. Rossi\inst{4},
   G. Ghirlanda\inst{1},
   and N. Schartel\inst{5}}

   \offprints{A.Tiengo, email: tiengo@mi.iasf.cnr.it}

   \institute{Istituto di Astrofisica Spaziale e Fisica Cosmica -- CNR,
              Sezione di Milano ``G.Occhialini'',
          Via Bassini 15, I-20133 Milano, Italy
         \and
             Universit\`{a} degli Studi di Milano,
            Dipartimento di Fisica, v. Celoria 16, I-20133 Milano, Italy
         \and
         INAF-Osservatorio Astronomico di Brera, v. Bianchi 46, I-23907 Merate (LC), Italy
         \and
         Institute of Astronomy, Madingley Road, Cambridge CB3 OHA, UK
         \and
          XMM-Newton Science Operation Center, ESA, Vilspa, Apartado 50727, 28080 Madrid, Spain   }

%   \date{Received September 15, 1996; accepted March 16, 1997}

\abstract{
We report on \xmm and \xte observations of the bright (fluence $\sim$ 10$^{-4}$ erg cm$^{-2}$)
and nearby (z=0.1685) Gamma-Ray Burst \gr associated to SN2003dh.
The first \xte observation, 5 hours after the burst,  shows  a flux
% 2-10 keV flux
%of the order of 10$^{-10}$ erg cm$^{-2}$ s$^{-1}$,
decreasing with time as a power law with index 0.9$\pm$0.3.
Such a decay law is only marginally consistent with a further \xte measurement
(at t-t$_{GRB}\sim$30 hr).
% and it largely overpredicts the flux at the
%time of the \xmm observation (t-t$_{\rm GRB}\sim$37 days).
Late time observations of this bright afterglow at X--ray wavelengths  have the
advantage, compared to optical observations, of not
being affected by contributions from the supernova and host galaxy.
A first  \xmm observation, at t-t$_{\rm GRB}\sim$37 days, shows a flux
of 4$\times10^{-14}$ erg cm$^{-2}$ s$^{-1}$ (0.2-10 keV). The spectrum is a power law
with photon index $\Gamma$=1.9 and absorption $<$2.5$\times10^{20}$ cm$^{-2}$,
consistent with the Galactic value. A further \xmm pointing at
t-t$_{\rm GRB}\sim$61 days shows a flux fainter by a factor $\sim$2.
The combined \xte and \xmm measurements require a break at t$\sim$0.5 days in the
afterglow decay, with a power law index increasing from 0.9 to 1.9, similar to
what is observed in the early part of the optical afterglow.
The extrapolation of the XMM-Newton spectra to optical frequencies lies a
factor of $\sim10$ below simultaneous measurements. This is likely due to
the presence of SN2003dh.
}

\authorrunning{A.Tiengo}

\maketitle

%
%________________________________________________________________

\section{Introduction}

A very  bright Gamma-Ray Burst (GRB) has been recorded by several satellites on
March 29, 2003.
The accurate localization obtained with HETE-2 after about 1 hour (\cite{vanderspek,ricker})
prompted rapid   observations that  revealed a bright
optical transient with R magnitude about 13 (\cite{peterson,torii}).
A redshift of z=0.1685 has been measured for the \gr host galaxy
(\cite{greiner,caldwell}).
This is the second smallest redshift determined for a GRB (the smallest one
is that of GRB980425,
likely associated with  SN1998bw at z=0.0085
(\cite{galama}) as supported by recent observations (\cite{pian})).
Although the intrinsic luminosity of \gr was in the low end of the distribution
for GRBs (see below), its proximity led to a very high fluence for the prompt emission
(1.2$\times10^{-4}$ erg cm$^{-2}$, 30-400 keV, \cite{ricker}) and to a bright afterglow which can
be studied with unprecedented detail at all wavelengths and over long timescales
after the burst explosion.
Among the most interesting results reported so far, is the finding of clear spectroscopic
signatures of an underlying supernova
(Stanek et al. (2003a), Hjorth et al. (2003)),
which provides strong evidence for the association between long GRBs and core collapse
supernovae.

The early phases of the X--ray afterglow of \gr  were observed with two \xte pointings
obtained 5  hours and 1.24  days after the burst
(\cite{marshall,marshall2}).
Unfortunately, no further X--ray data could be collected during the following month.
At the beginning of May the GRB position became compatible with the
visibility constraints of the \xmm satellite, which performed two observations
37 days and 61 days after the burst.
The brightness of \gr, coupled with the large collecting area
of \xmm, allow us to study in detail for the first time an X--ray afterglow at
such long times after the prompt emission.
This is particularly interesting since at these wavelengths we do
not expect significant contamination from the underlying supernova and/or host galaxy.

\section{Data analysis and results}

\subsection{RossiXTE}

The first \xte observation,  consisting of two time intervals
of duration 1500 s and 500 s, respectively, was done on March 29, about 5 hours
after the GRB.
The Proportional Counter Array  instrument (PCA, Jahoda et al. 1996)
aboard \xte consists of five Proportional Counter Units (PCU).
Only three (n. 0,2 and 3) and two (n.0 and 2) of them
were on during the first and second time interval, respectively.
Since the instrumental gain is known to vary between PCUs, we separately extracted
the spectra from the two time intervals. In order to increase the signal to noise ratio,
only the top layer anodes were used in the analysis.
The corresponding response matrices were generated using PCARSP V8.0 and
the background spectra were estimated with the faint-source model as input to
PCABACKEST V3.0.
All the errors quoted below are at the 90\% confidence level.

\begin{figure}
      \vspace{2cm}
      \hspace{0cm}\psfig{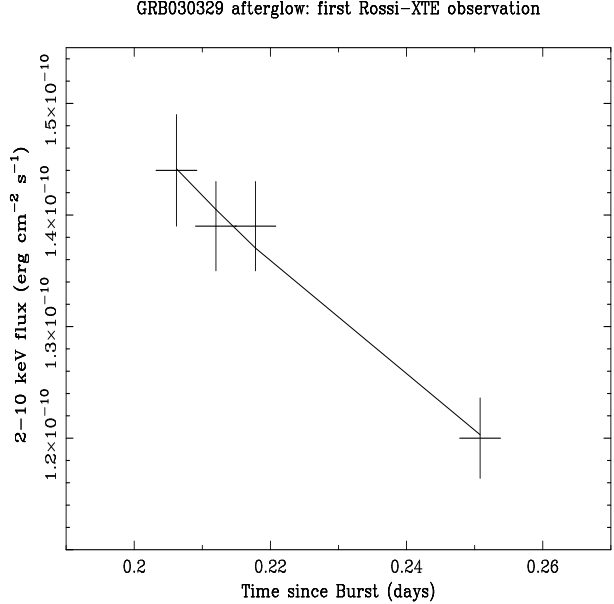}
      \caption[]{X--ray afterglow of \gr as measured during the first
      \xte observation. Each time bin is 500 s long.
            The line is the best fit with a power law of index  $\delta$=0.9$\pm$0.3.}
         \label{xte}
   \end{figure}

Since the source spectral shape did not vary between the two intervals
(except for the normalization), we  fitted   them together,  obtaining a
best fit with a power law model with photon index
$\Gamma$=2.17$_{-0.03}^{+0.04}$ and absorption
N$_{\rm H}<$5$\times10^{21}$ cm$^{-2}$.
The average flux in the first interval was
F$_{\rm x}$=(1.38$_{-0.02}^{+0.05}$)$\times10^{-10}$ erg cm$^{-2}$ s$^{-1}$
%, while it
%was F$_{x}$=(1.18$\pm$)$\times10^{-10}$ erg cm$^{-2}$ s$^{-1}$
(2-10 keV).
%
%
%In the second part of the observation the flux is
%F$_{x}$=(1.18$\pm$)$\times10^{-10}$ erg cm$^{-2}$ s$^{-1}$  (2-10 keV).
During this observation the flux decreased with time as a power law with
index $\delta$=0.9$\pm$0.3 (see Fig.  \ref{xte}).

A second observation was performed 30 hours after the GRB,
on March 30 from 17:30 to 19 UT, with a net observing time of 3500 s.
Due to the lower flux, the derived spectral parameters are less constrained.
A power law fit yields $\Gamma$=1.8$_{-0.2}^{+0.3}$,
N$_{\rm H}<$3.7$\times10^{22}$ cm$^{-2}$, and
F$_{\rm x}$=(1$_{-0.1}^{+0.2})\times10^{-11}$ erg cm$^{-2}$ s$^{-1}$  (2-10 keV).
%
%As can be seen in Fig. \ref{xmm_lc}, t
This flux is smaller than the extrapolation of the power
law decay found in the first observation, suggesting the presence of a break in the
light curve.
A single power law fit to all the \xte data gives only a marginally acceptable fit
for a slope  $\delta\sim$1.5 ($\chi^{2}$=7.6 for 3 degrees of
freedom (dof)).

Finally, we analyzed an observation carried out on April 6.
The afterglow was not detected, with a flux upper limit
of 2.6$\times10^{-12}$ erg cm$^{-2}$ s$^{-1}$ (3$\sigma$).

\subsection{XMM-Newton}

\xmm observed the position of \gr starting on May 5, 2003 at 12:30 UT,
for an observation length of  $\sim$12 hours.
The last part of the observation was affected by high
particle background and was therefore excluded from our analysis,
resulting in net exposure times of 29 and 32 ks, respectively
in the PN and MOS cameras
of the EPIC instrument (\cite{struder,turner}).
All the cameras operated
in Full Frame mode and with the thin optical blocking filter.
The data were processed using SAS version 5.4.1.

\begin{figure}
      \vspace{2cm}
      \hspace{0cm}\psfig{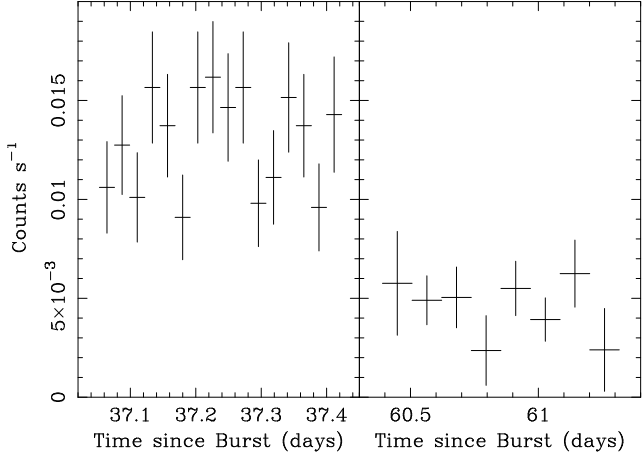}
     \caption[]{Background subtracted
      PN light curve of the X--ray afterglow during the first (left) and second (right)
      \xmm observation.
      The bin sizes are  2,000s and   10,000   s, respectively. Note that the count rate quoted in the text is higher
      than the one shown here since it has been corrected for the fraction of source
      counts falling outside the extraction region.}
         \label{lcurve}
   \end{figure}

A  source with a PN net count rate of 0.018 counts s$^{-1}$  was  detected
at R.A. = 10$^h$ 44$^m$ 49.9$^s$,
Dec. =  +21$^{\circ}$ 31$'$  15$''$ (J2000, error radius of 4$''$),
consistent with the position of \gr.
Its flux during the observation is consistent with a constant value
(see Fig.  \ref{lcurve}).
A second source of similar intensity (0.022  PN counts s$^{-1}$)
is present at a distance of $\sim$30$''$ to the NW. Its spectrum
(a power law with photon index $\Gamma$=1.7$\pm$0.2 and
N$_{\rm H}$=(5$\pm$2)$\times10^{21}$ cm$^{-2}$)
and its positional coincidence with a galaxy at z=0.136
(\cite{krisciunas}) indicate that  this source is an AGN.

To measure the afterglow spectrum we used an extraction radius of 15$''$ in
order to minimize the contamination from the AGN. For the
extraction of the background spectrum we chose a circular region
(radius 15$''$) at the same distance from the AGN as the afterglow.
This was done to take into account the small contamination from the AGN
to the spectrum
(we estimate that at most 20\% of the counts
in the source extraction region could be due to the AGN).
The spectra, over the 0.2--10  keV energy range,
were rebinned in order to have at least 30 counts per channel.
After checking that consistent results were obtained in the three cameras, we fitted
jointly the MOS and PN data.
%All the errors quoted below are at the 90\% confidence level.

\begin{figure}
      \vspace{2cm}
      \hspace{0cm}\psfig{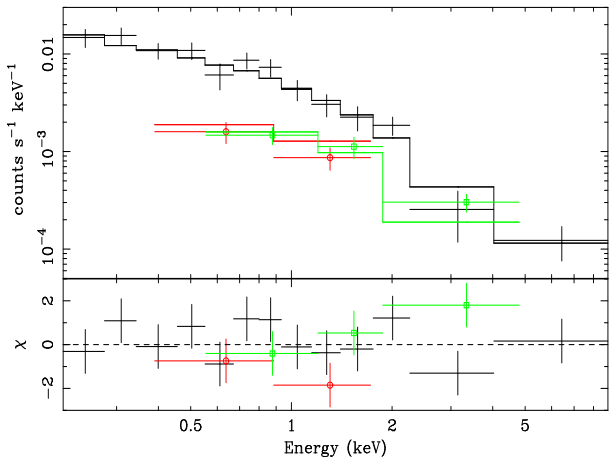}
      \caption[]{EPIC spectrum of the X--ray afterglow of \gr fitted with a power
        law model. Upper line and data refer to the PN, lower ones to the MOS.
         The bottom panel shows the best fit residuals in units of standard deviations. }
       \label{spectrum}
   \end{figure}

The best fit ($\chi^{2}/dof$=16.5/15, see Fig.  \ref{spectrum})
was obtained with an absorbed  power law
with photon index $\Gamma$ = 1.92$_{-0.15}^{+0.26}$,
N$_{\rm H}<$2.5$\times10^{20}$ cm$^{-2}$,
and flux F$_{\rm x}$=(4.2$\pm$0.5)$\times10^{-14}$ erg cm$^{-2}$ s$^{-1}$  (0.2-10 keV).
The Galactic absorption in this direction is  N$_{\rm H}$=2$\times$10$^{20}$ cm$^{-2}$
(\cite{dickey}).
Other simple models, e.g. blackbody, thermal bremsstrahlung, thermal plasma (MEKAL)
gave unacceptable fits (the respective values of $\chi^{2}$/dof being 73.5/15,
26.3/15 and 36.8/15).

% Comparison with the   published X--ray fluxes
% (\cite{marshall,marshall2}), indicated
% an afterglow  decreasing in time as a
% power law  with index  $\delta\sim$1.7 (\cite{tiengo}).
% However, our  more detailed analysis
% of the \xte data, presented in the next section, gives evidence for
% a more complex time evolution.

A second \xmm pointing  started
on May 28, 2003 at 21:00 UT. It lasted about one day, but it was severely
affected by periods of  high particle background, which were excluded
in our analysis, resulting in about 40 ks of useful data.
The analysis was performed as described above for the first observation.
The source at the \gr position fainted to
F$_{\rm x}$=(2.2$_{-0.3}^{+0.5}$)$\times10^{-14}$ erg cm$^{-2}$ s$^{-1}$  (0.2-10 keV),
thus confirming that it is indeed the GRB afterglow.
Its spectrum could be fit by an absorbed  power law with $\Gamma$ = 2.1$_{-0.2}^{+0.4}$
and N$_{\rm H}<$4.5$\times10^{20}$ cm$^{-2}$ ($\chi^{2}$/dof=4.5/13).
Also in this case a thermal model gave a worse fit ($\chi^{2}$/dof=21.3/13 for a MEKAL).

\section{Discussion}

\begin{figure}
\psfig{figure=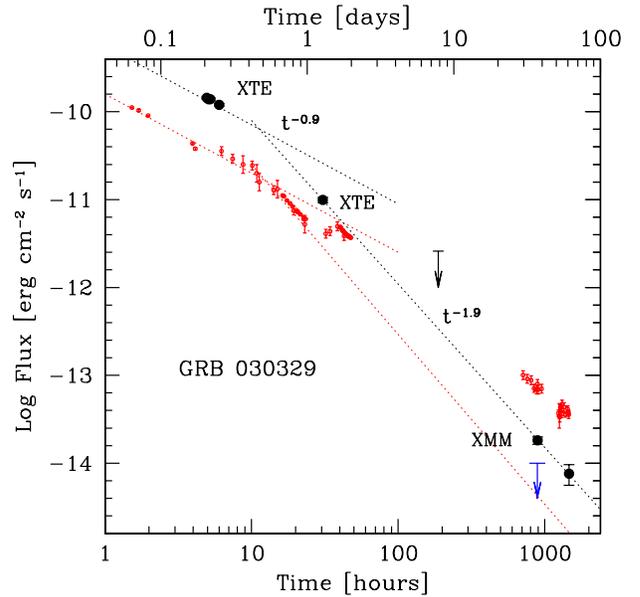,angle=0,width=9cm}
\vskip -0.5 true cm
\caption{The X--ray light curve of GRB030329 (large dots and upper limit
at $\sim$8 days) is compared
with the optical data at times close to the epochs of the X--ray
observations.
Optical data are from
Burenin et al. (2003a);
Burenin et al.  (2003b);
Fitzgerald \& Orosz (2003);
Ibrahimov et al. (2003);
Price \&  Mattei (2003);
Price (2003);
Rykoff \& Smith (2003);
Stanek et al. (2003b);
Stanek et al. (2003c);
Zharikov et al. (2003).
The upper limit at 37 days corresponds to the
optical flux calculated through the extrapolation of the \xmm
spetrum (see Fig. \ref{sed}).
The dotted lines are only indicative of the time decay slopes,
and are not fits to the optical data.
}
\label{xmm_lc}
\end{figure}

All our  measurements of the X--ray afterglow of \gr for the 2-10 keV
range are plotted in Fig. \ref{xmm_lc},
%(the \xmm fluxes derived above have been converted to
%the 2-10 keV used for \xte).
where one can  see that the  break in the afterglow decay suggested by the second \xte
observation is clearly confirmed by \xmm.
While a power law index  $\delta$=0.9 was found during the first \xte observation,
a fit to the following points gives a slope $\delta$=1.86$\pm$0.06.
% also plotted in Fig. \ref{xmm_lc}.
We   estimate that the break occurred in the time interval
0.3 - 0.8 days, with a most likely value of t$_{\rm break}$=0.45 days.
This value is consistent with the time of the break in the optical afterglow
(t$_{\rm break}$=0.48 days, Price et al. 2003).
%(see the optical data plotted in Fig. \ref{xmm_lc}).

This achromatic break can be readily explained as a
``jet--break", due to the decreasing bulk Lorentz factor
$\Gamma$, making $1/\Gamma$ equal to the jet opening
angle (see Rhoads 1999).
Following Frail et al. (2001) it is then possible to
estimate the opening angle of the jet ($\sim3^{\circ}$),
corresponding to $\Gamma\sim 19$ at the time
of the break.
With this opening angle, the ``true" energy radiated by the
burst in $\gamma$--rays turns out to be
$E_\gamma= 3\times 10^{49}$ erg (see Frail et al. 2001 for the
relevant uncertainties concerning these estimates).
This value is
at the very low end of the distribution found by Frail et al. (2001),
making GRB030329 an atypically weak burst.
Note also that the optical light curve shows several achromatic
``rebrightnenings" and breaks (Granot et al. 2003, and references therein),
which weaken the association
of the first break with the jet--break.
The paucity of the X--ray data does not allow us to infer if the X--ray
follows the optical during the several rebrightenings occurring in
the optical band, which could help to investigate the origin
of such rebrightenings (Lazzati et al. 2002; Granot et al. 2003).

\begin{figure}
\psfig{figure=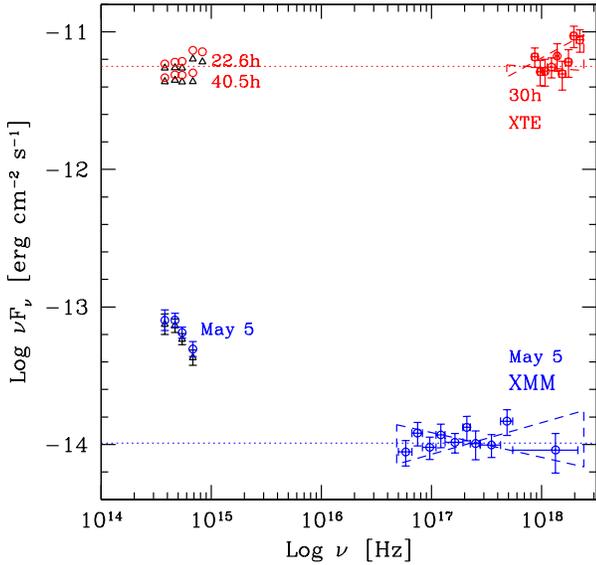,angle=0,width=9cm}
\vskip -0.5 true cm
\caption{The quasi simultaneous optical to X--ray SED of \gr
at the epochs of the second \xte and of the first \xmm observation.
Optical data are from Zharikov et al. (2003),
Fitzgerald \& Orosz (2003) and Ibrahimov et al. (2003).
Open circles refer to fluxes de--reddened assuming an extinction of
$A_V=0.16$ consistent with both the Galactic value of the column density and
 the N$_H$ found for the XMM-Newton fit. Triangles assume instead $A_V=0$.
For the SED at $\sim 30$ hours, the optical and X--ray data
lays on the same power law [$F(\nu)\propto \nu^{-1}$, top dotted line].
The bottom dotted line corresponds to the same spectral shape for
the May 5th SED.
}
\label{sed}
\end{figure}

In Fig. \ref{sed} we show the simultaneous optical and X--ray spectra
corresponding to the second \xte  and to the first \xmm
observation epochs.
Since we did not find an exactly simultaneous
spectrum at $\sim 30$ hours after the trigger, we show in
Fig. \ref{sed} the optical points taken immediately before (22 hours)
and after (40 hours) the \xte  observations.
As can be seen, at this epoch the extrapolation of the
optical spectrum joins very smoothly the X--ray data, and fits both
their normalization and slope.
Optical and X--ray fluxes therefore
belong to the same spectral segment characterized by
$F(\nu)\propto \nu^{-1}$ (i.e. a flat spectrum in $\nu F(\nu)$).
This shape can be explained by the standard synchrotron--external
shock model (e.g. Sari et al. 1998)
as due to a population of
relativistic electrons injected in the emitting region with an energy
distribution $\propto \gamma^{-p}$ with $p\sim 2$,
with both the optical and X--ray frequencies laying
(at $\sim$30 hr) above the cooling frequency $\nu_c$
(this is the frequency produced by those electrons that have
just cooled in a dynamical time).

The optical and X--ray decay slopes
before and immediately after the break at $\sim 0.5$ days
are the same (i.e. $F(t) \propto t^{-0.9}$ and $F(t) \propto t^{-1.9}$).
The first decay slope is consistent with what expected in the case
of $p\sim 2$, slow cooling regime and the cooling frequency
below the optical (Panaitescu \& Kumar, 2000).
In this case the decay index is independent on the density profile.
The second decay slope is instead consistent
with what expected after the jet break ($t^{-p}$) if the
jet matter expands laterally at a velocity close to the
speed of light (Rhoads 1999).

The evolution of the high-energy spectrum depends on the behavior in time
of the cooling frequency, which in turn depends on the
circumburst density profile and on the jet dynamics.
Therefore, after 30 hours, there are two possibilities.
%If the jet undergoes sideway expansion at 0.5 days
If the circumburst medium is  homogeneous, the cooling frequency $\nu_{\rm c}$
decreases as $t^{-1/2}$ (before the jet break) or remains constant
(after the jet break with sideway expansion), leaving the optical to X--ray slope unaltered.
Instead, in the case of a $R^{-2}$ wind density profile, the corresponding behaviours
are $\nu_{\rm c} \propto t^{1/2}$ and $\nu_{\rm c} $=const.
Only in the wind case before the jet break $\nu_{\rm c}$
possibly overtakes the optical band
(see e.g. Panaitescu \& Kumar 2000). This gives  a spectral break between the optical and
the X--ray bands, accompanied by a {\it flattening} of the optical
light curve (due to the appearance of the $\nu<\nu_{\rm c}$ harder spectral
slope in the optical band).
We conclude that the optical to X--ray flux ratio of
the synchrotron--external shock component either remains fixed
 or decreases in time.

The optical data of May 5th define a steep (spectral index
$\alpha_{\rm opt}>1$) spectrum and lay above the extrapolation of
the X--ray spectrum.
We interpret this as evidence for a significant contribution,
in the optical, of SN2003dh.
The X--ray flux can be used to estimate an upper limit to the flux of the
optical synchrotron--external shock component (indicated by the dotted
line in Fig. \ref{sed}): $\nu F(\nu) < 10^{-14}$ erg cm$^{-2}$ s$^{-1}$.
It is an upper limit if $\nu_{\rm c}>\nu_{\rm opt}$; in the case of
homogeneous circumburst density this becomes the actual value.
We conclude that the optical lightcurve, around May 5th,
is dominated by the light from the supernova SN2003dh,
by at least two magnitudes (see also Fig. \ref{xmm_lc}).

Similar considerations apply to the data at t-t$_{\rm GRB}$= 61 days,
although with larger uncertainties owing to the less constrained \xmm
spectral slope.

%

%\subsection{Jet break}

%As mentioned, the X--ray light curve shows a break at approximately
%0.5 days after the trigger, which is consistent with the break time of
%the optical light curve (see Fig. \ref{xmm_lc}).

\section{Summary}

Thanks to the high sensitivity of \xmm we could study  the
optical--X--ray SED
of the afterglow of \gr  and its time evolution up to late times.
%In fact the high sensitivity of \xmm
%allow  to detect
%very small X--ray flux values and to explore the behavior of the
%high-energy afterglow emission at late times.
This is particularly
important for this burst due to its association with the
supernova 2003dh, which, at late times,  contributes
to the optical flux (Stanek et al. (2003a), Hjorth et al. (2003)).
%We show in the following that t
The early and late time X--ray data, combined with the simultaneous optical
detections, have been used to estimate the contributions of the
(non--thermal) afterglow and supernova components at optical
frequencies.

Our main results are the following:

%Our analysis  of the \xmm and \xte observations of the bright X--ray
%afterglow of \gr, interpreted in the context of the available information
%on the optical afterglow, allows us to derive the following results:

\begin{itemize}

\item The first epoch \xte data define an X--ray light curve
decaying in time as $t^{-0.9}$. This decay index is consistent with  the
one of the optical flux.

\item The two \xmm observations at late epochs    yield spectra
well fit by a power law with photon index $\sim$2,

\item The \xte and \xmm  data, taken together,
are consistent with a break in the light curve occurring at
$\sim$0.5 days, simultaneously with the optical break. After this break
the afterglow decays as $t^{-1.9}$.

\item The optical to X--ray SED at 30 hours strongly indicates
that both spectral bands lay on the same branch, above
the cooling frequency.

\item The optical to X--ray SED on May 5$^{th}$ (and possibly also on May 28$^{th}$)
indicates instead an
optical excess that we interpret as due to SN 2003dh,
which should dominate (by a factor $\sim$10) the non--thermal
optical emission.

\end{itemize}

\begin{acknowledgements}
Based on observations obtained with XMM-Newton, an ESA science
mission  with  instruments and contributions directly funded by
ESA Member States and NASA.
We are grateful to the \xmm Project Scientist Fred Jansen for granting time to observe this source.  This
research has made use of the data and resources obtained through the HEASARC on-line service, provided by NASA Goddard Space Flight
Center.
This work has been supported by the Italian Space Agency.
ER thanks the Brera Observatory for hospitality during
the completion of this work.
GG acknowledges the MIUR for the COFIN grant.
\end{acknowledgements}

\end{document}